\documentstyle[aps,prl,twocolumn,psfig,floats]{revtex}
\newcommand{\be}{\begin{eqnarray}}
\newcommand{\ee}{\end{eqnarray}}
\newcommand\del{\partial}\begin{document}
\draft
\wideabs{
\title{ Universality in Chiral Random Matrix Theory at $\beta =1$
and $\beta =4$}
\author{M.K. \c Sener and J.J.M. Verbaarschot}
\address{Department of Physics and Astronomy\\
SUNY, Stony Brook, New York 11794}
\date{\today}
\maketitle
\begin{abstract}
In this paper the kernel for the spectral correlation functions of the 
invariant chiral random matrix ensembles with real ($\beta =1$) and 
quaternion real ($\beta = 4$)
matrix elements is expressed in terms of the kernel of the corresponding
complex Hermitean random matrix ensembles ($\beta=2$). 
Such identities are exact
in case of a Gaussian probability distribution and, under certain smoothness
assumptions, they are shown to be valid 
asymptotically for an arbitrary finite polynomial potential.  
They are proved by means of
a construction proposed by Br\'ezin and Neuberger. 
Universal behavior at the hard edge of the spectrum for all three chiral 
ensembles then follows from microscopic universality for $\beta =2$ 
as shown by Akemann, Damgaard, Magnea and Nishigaki.
\end{abstract}
\pacs{PACS numbers: 64.60.Cn, 05.45.+b, 11.30.Rd, 12.38.Aw}
}

\narrowtext 
Since its introduction in nuclear physics \cite{Porter}, Random Matrix Theory
(RMT) has been applied successfully to many different branches of physics 
ranging from atomic physics to quantum gravity (for a recent comprehensive
review we refer to \cite{hdgang}). One important common ingredient is that 
eigenvalue correlations appear to be insensitive to the details of the 
underlying Hamiltonian. The success of RMT is based on this type of 
universality, and it is no surprise that it has received a great deal of 
attention in recent literature 
\cite{AJM,Been,bzgeneral,Hack,Itoi,Nishigaki-uni,Brezinn,brezin-hikami-zee}
\cite{Pzinn,Zee-poly,Deo,GWu,Sener1,Bowick,KF,kanzieper,Eynard,Brezin-Hikami}
\cite{Beenakker,andreev,Akemann,Akemannmulti,Damgaardmass,wilke,Damgaard}.
What has been shown is that spectral correlators on the scale of the average
eigenvalue spacing are insensitive to the details of the probability 
distribution of the matrix elements. Because of its mathematical simplicity
most studies were performed for complex ($\beta = 2$) Hermitean RMT's. However,
in the case of the classical RMT's it was shown that 
universality extends to real ($\beta = 1$) and quaternion real ($\beta =4$)
matrix ensembles \cite{Been,Hack,Itoi}. 
This suggests that relations between
correlation functions for different values of $\beta$ which can be derived
for a Gaussian probability distribution \cite{V2,Berbenni} might be valid
for  a wide class of probability distributions. The main goal of this paper
is to establish such general relations. As a consequence, universality for
the much simpler complex ensembles implies 
universality for the real and quaternion real ensembles.

In this letter we address the question of microscopic 
universality for the chiral 
ensembles. These ensembles are relevant for the description of spectral
correlations of the QCD Dirac operator. They also appear in theory of universal
conductance fluctuations in mesoscopic systems \cite{Slevin,martin}.
In particular, they can be applied to spectral correlations near
$\lambda = 0$. According to the Banks-Casher formula \cite{BC}, this part
of the spectrum is directly related to the order parameter $\Sigma$ of the 
chiral phase transition ($\Sigma = \lim \pi \rho(0)/V$, where $V$ is the
volume of space time and $\rho(\lambda)= \sum_k \delta(\lambda-\lambda_k) $).
It is therefore natural to introduce the microscopic limit where the variable
$u = \lambda V \Sigma$ is kept fixed for $V\rightarrow \infty$. 
For example, the microscopic spectral density is defined by \cite{SV}
\be
\rho_S(u) = \lim_{V\rightarrow \infty} \frac 1{V\Sigma} \langle
\rho (\frac u{V\Sigma}) \rangle,
\label{BC}
\ee
where the average is over the distribution of the matrix elements of the
Dirac operator.
Successful applications of the chiral ensembles to lattice 
QCD spectra can be found in \cite{Berbenni,Ma,Markum,HV}.

The chiral random matrix ensembles for $N_f$ massless quarks in the sector
of topological charge $\nu$ are defined by
the partition function \cite{SV,V}
\be
Z_{N_f,\nu}^\beta =
\int DW  {\det}^{N_f}\left (\begin{array}{cc} 0 & iW\\
iW^\dagger & 0 \end{array} \right )
e^{-{n \beta} {\rm Tr}V(W^\dagger W)},
\label{zrandom}
\ee
where $W$ is a $2n\times (2n+\nu)$ matrix.
As is the case in QCD, we assume that 
$\nu$ does not exceed $\sqrt {2n}$. The parameter $2n$
is identified as the dimensionless volume of space time.
The matrix elements of $W$ are either real ($\beta = 1$, chiral
Orthogonal Ensemble (chOE)), complex
($\beta = 2$, chiral Unitary Ensemble (chUE)),
or quaternion real ($\beta = 4$, chiral Symplectic Ensemble (chSE)).
For technical reasons we only consider finite polynomials potentials 
$V(x)$. The simplest case is the Gaussian case with $V(x) = \Sigma^2 x$
(also known as the Laguerre ensemble).

It was shown by Akemann et al. \cite{Akemann} that, for $\beta =2$,
the microscopic spectral density and the 
microscopic spectral correlators do not depend on the potential $V(x)$
and is given by the result \cite{VZ} for the Laguerre ensemble. 
For $\beta = 2$ all spectral correlators can be obtained from  
an orthogonal polynomial kernel corresponding to 
the probability distribution. In their proof the
Christoffel-Darboux formula is used to express this 
kernel in terms of large order polynomials. Microscopic 
universality then follows from the 
asymptotics of orthogonal polynomials. As a remarkable achievement, they
were able to generalize the relation for Laguerre polynomials
\be
\lim_{n\rightarrow \infty} n^{-a}L_n^a(\frac xn) = x^{-a/2}J_a(2\sqrt x)
\label{Laguerre}
\ee
to orthogonal polynomials corresponding to an arbitrary polynomial potential. 
However, their work cannot easily be generalized to $\beta = 1$ and $\beta = 
4$. The main result of the present work is 
a relation between the kernels
for the correlation functions of the chOE and chSE and the kernel
of the chUE. This relation is exact for the Gaussian ensembles and is valid
asymptotically for an arbitrary polynomial potential. For $\beta=4$, this
relation shows universality of the microscopic spectral density 
and correlators (for
$\beta=1$ only a partial proof was obtained).

The  partition function (\ref{zrandom}) is invariant under
$ W \rightarrow U^\dagger W V$ 
where the matrices $U$ and $V$ with dimensions determined by $W$
are orthogonal for $\beta=1$, unitary 
for $\beta = 2$, and symplectic for $\beta = 4$. This invariance
makes it possible to express the partition function (\ref{zrandom}) 
in terms of the eigenvalues $x_k$ of $W W^\dagger$ as
\be
Z_{N_f, \nu}^{\beta} =
\int \prod_k dx_k x_k^{2a} |\Delta(x_i)|^\beta
 e^{-n\beta \sum_k V(x_k)},
\label{zeig}
%%correction: eliminated 2N_f
\ee
where the Vandermonde determinant is defined by
$\Delta(x_i) = \prod_{k<l} (x_k-x_l)$
and $a = N_f-1+\beta\nu/2 +\beta/2$. 

For $\beta = 2$, the spectral correlation functions can be evaluated 
\cite{mahoux} by
expressing the Vandermonde determinant in terms of the orthogonal polynomials
defined by
\be
\int_0^\infty dx e^{-2\phi_a(x)} q^{2a}_k(x)q^{2a}_l(x) = \delta_{kl},
\label{ortho}
\ee
where we have introduced the potential $\phi_a(x) = nV(x) - a\log x$.
By using orthogonality relations it can be shown that all spectral 
correlation functions can be expressed in terms of the kernel
\be
K_{2n}^{2a}(x,y) = \sum_{k=0}^{2n-1} q^{2a}_k(x)q^{2a}_k(y).
\ee
The spectral density is given by $K_{2n}^{2a}(x,x)\exp(-2\phi_a(x))$. 
Microscopic
universality then follows from the following generalization of (\ref{Laguerre})
\cite{Akemann}  
\be
\lim_{n\rightarrow\infty} \left . \sqrt{h_k^{2a}} 
q_k^{2a}(\frac {x^2}{n^2}, n)\right
|_{k=tn} = \Gamma(2a+1) \frac{J_{2a}(u(t)x)}{(u(t)x/2)^{2a}},
\label{Ake}
\ee
in the normalization $q_k^{2a}(0) \sqrt{h_k^{2a}} = 1$. The function $u(t)$ 
follows from the asymptotic properties of the leading order coefficients of
the $q_k^{2a}(x)$ and the normalizations $h_k^{2a}$. Its value at
$t=1$ is given by $u(1) = 2\pi \rho(0)$.
     
In order to perform the integrations by means of orthogonality relations
for $\beta = 1$ and $\beta = 4$, one has to introduce the skew-orthogonal 
polynomials \cite{Dyson_skew,mahoux}. Below, we first discuss the case
$\beta = 1$ and then give general outlines for the case $\beta = 4$.

For $\beta = 1$, the skew orthogonal polynomials of the second kind are
defined by
\be
<R_i,R_j>_R = J_{ij}.
\label{orhtoR}
\ee
with the skew orthogonal scalar product
\be
<f,g>_R = \int_0^\infty dx e^{-2\phi_a(x)} f(x) \hat Z g(x),
\ee
and nonzero matrix elements of  $J_{ij}$ given by
$J_{2k, 2k+1}= -J_{2k+1,2k} = -1$.
The operator $\hat Z$ is defined by
\be
\hat Z g(x) = \int_0^\infty dy e^{\phi_a(x)} \epsilon(x-y) e^{-\phi_a(y)} g(y),
\label{Z}
\ee
Here, $\epsilon(x) = x/2|x|$.
It can be shown that all correlation functions can be expressed
in terms of the kernel \cite{Dyson_skew,mahoux}
\be
K_1(x,y) = \int_0^x dz e^{-\phi_a(z)} 
 k_1(z,y)e^{-\phi_a(y)},
\label{K1def}
\ee
where we have introduced the pre-kernel
\be
k_1(y,z) = \sum_{i,j=0}^{2n-1} R_i(y) J_{ij} R_j(z).
\label{prekernel}
\ee
In particular, the spectral density is given by
\be
\rho(x) = K_1(x,x)-\frac 12 K_1(\infty,x).
\label{rho}
\ee

A general scheme for the construction of skew-orthogonal polynomials was 
introduced by Br\'ezin and Neuberger\cite{BN}.  The idea is to express 
them in terms of orthogonal polynomials defined by (\ref{ortho}). 
For technical reasons we expand in the polynomials $q_k^{2a+1}$
(with weight function $x^{2a+1} \exp(-2nV(x))$). 
The skew-orthogonal polynomials of degree $i$ can thus be expressed as
\be
R_i(x) = \sum_{j=0}^i T_{ij} q_j^{2a+1}(x),
\ee
where $T$ is a lower triangular matrix with nonvanishing diagonal elements. 
An essential role is
played by the inverse, $\hat L$, of the operator $\hat X^{-1} \hat Z$ with
$\hat Z$ defined
in (\ref{Z}) and $\hat X g(x) =  xg(x)$.
It can be easily verified that 
\be
\hat L = \hat X (\hat \del - \phi_a'(\hat X)) + \hat 1.
\label{L}
\ee
The matrix representations of the operators $\hat X$, $\hat X \hat \del$, 
$\hat X^{-1} \hat Z$ and $\hat L$ in the basis $q_k^{2a+1}$
will be denoted by $X_{kl}$, $D_{kl}$,
$Y_{kl}$ and $L_{kl}$, respectively. 
In the remainder of this derivation the index $2a+1$ will be suppressed.

In matrix notation (\ref{orhtoR}) can be rewritten as
\be
T Y T^T = -J.
\ee
By using that $LY = 1$, this relation can be expressed as
\be
L =  T^T J T.
\label{ojo}
\ee
It can be shown that the matrix $L_{kl}$ is a band matrix with width determined
by the order of the polynomial potential $V(x)$. It then follows that $T$ is
a band matrix as well \cite{BN}. 
For example, for a Gaussian potential we have that
$T_{2m,k} = a_0 \delta_{2m,k}$ and $T_{2m+1,k} = b_0 \delta_{2m+1,k}
+b_1 \delta_{2m,k} + b_2 \delta_{2m-1,k}$ with coefficients derived in 
\cite{Nagao,V2}.

It turns out that we do not need explicit expressions for the $T_{ij}$. The
pre-kernel (\ref{prekernel}) can be expressed as
\be
k_1(x,y) = \sum_{i,j=0}^{2n-1} \sum_{k\le i}\sum_{l\le j}q_k(x) 
T^T_{ki} J_{ij} T_{jl} q_l(y).
\label{tjt}
\ee
In this relation the indices $i$ and $j$ run up to $2n-1$ in contradistinction
to the relations (\ref{ojo}) where they run up to $\infty$. However,
it follows from the band structure of $L$ that the number of terms 
outside the range in (\ref{tjt}) is of the same order as 
the degree of the polynomial potential which is finite. 
These terms are negligible in the continuum limit of the type (\ref{Ake})
where the $q_k^{2a+1}(x)$ and $L_{kl}$ depend smoothly on $k$ and $l$ (notice
that $L_{kl}$ is not smooth in $|k-l|$). However, for $x$ around zero and
$y$ near the largest zero of $q_l(y)$ we expect potentially non-negligible 
contributions.
We thus have that
\be
k_1(x,y) \simeq \sum_{k,l=0}^{2n-1}  q_k(x) L_{kl} q_l(y).
\label{k1fin}
\ee
By means of a partial integration
the matrix elements of $\hat 1-\hat X\phi_a'(\hat X)$ in $L$ 
can be expressed in 
terms of the matrix 
elements of $\hat X \hat \del$.  This results in
\be
L_{kl} &=& \frac 12 \int_0^\infty zdz e^{-2\phi_a(z)} 
(q_l(z) z\del q_k(z) - q_k(z) z\del q_l(z))\nonumber \\
&=&\frac 12 (D_{kl} -D_{lk}).
\ee
%After writing the derivative operators in terms of their 
%matrix elements we obtain for the pre-kernel  $k_1(x,y) $ 
%\be
%k_1(x,y) &\simeq& \frac 12 \int_0^\infty z dz e^{-2\phi_a(z)}
%\left ( q_k(x) q_l(z) D_{km} q_m(z)q_l(y) \right .\nonumber \\
%&-&\left . q_k(x)D_{lm}q_m(z) q_k(z) q_l(y)\right )
%\label{qqdqq}
%\ee
%The matrix $D$ is given by $D_{kl} = L_{kl} + 
%(\hat X\phi_a'(\hat X)-\hat 1)_{kl}$. 
%Since the matrix
%$L$ is anti-symmetric and the matrix $D$ is symmetric, we have
%$D_{kl} = - D_{lk} +2(\hat X \phi_a'(\hat X)-\hat 1)_{kl}$. 
%By substitution of  this relation
%in (\ref{qqdqq}), the transpose of $D$ can be re-expressed
%as $x\del_x$ or $y\del_y$.
%After performing the  $z$-integration by means of orthogonality,
The matrix elements of $D$ can be re-expressed
as $x\del_x$ or $y\del_y$. 
%After a cancellation due to the symmetry of 
%$(\hat X \phi_a'(\hat X)-\hat 1)_{kl}$
We finally arrive at a remarkably simple expression for $k_1(x,y)$,
\be
k_1(x,y) \simeq \frac 12(y\del_y -x\del_x) K_{2n}^{2a+1}(x,y).
\label{precentral}
\ee
With the help the asymptotic properties of the $q_k^{2a}$ (which are the
same as for the Laguerre polynomials) this relation
can be further simplified to (up to an overall factor determined by the
average spectral density)
\be
k_1(x,y) \sim \frac 12(\del_y -\del_x) K_{2n}^{2a}(x,y).
\label{central}
\ee
This is the central result of this paper. It is valid asymptotically 
both at the hard and the soft edge of the spectrum where a continuum
limit of the orthogonal polynomials $q_k^{2a+1}$ exists. 
However, as will be argued below, the result (\ref{central}) is not valid
for $x$ near the hard edge and $y$ at the soft edge of the spectrum.
This result relates the orthogonal pre-kernel to the 
unitary kernel  $K_{2n}^{2a}(x,y)$
which has been studied elaborately in the literature
\cite{Bronk,Kahn,Widom}. The relation (\ref{central}) is
exact for a Gaussian potential in which case it 
coincides with the result obtained
in \cite{V2,Forrester-Nagao,Nagao-Sl,brezin-hikami-zee,martin}.

Universality of the unitary kernel $K_{2n}^{2a}(x,y)$ 
at the hard edge has been well established
\cite{Akemann} for the chiral ensembles, whereas universality 
at the soft edge was shown in \cite{Bowick,KF}. We therefore expect
universal behavior of $k_1(x,y)$ in these domains. 

Let us finally focus on the spectral density. 
Using (\ref{rho}) and (\ref{central}), for a Gaussian potential it can be
expressed as
\be
\label{micro}
\rho(x) \simeq e^{-\phi_a(x)}\int_0^\infty dy e^{-\phi_a(y)} 
\epsilon(x-y) \frac 12
(\del_y -\del_x) K_{2n}^{2a}(x,y).
\nonumber\hspace*{-1cm}\\ 
\ee
In the microscopic limit where $n \rightarrow \infty$ at fixed $z\equiv x n^2$
the factor $\exp(-2nV(x)) \rightarrow 1$ and $K_{2n}^{2a}$ approaches its
universal limit. However, in one of the terms contributing to the integral 
the microscopic limit and the integration cannot be interchanged.
It can be shown that there is an additional contribution 
with $x$ near zero and $y$ near the edge of the spectrum. 
Naively taking into account this contribution  for nongaussian potentials
leads to a microscopic spectral density that differs from the universal 
expression. Alternatively, we have established universality of the
microscopic spectral density by means of Monte-Carlo simulations. Apparantly,
the smoothness assumptions in the  derivation of (\ref{precentral}) 
are violated in this case. 
The edge contribution resides in the term $K_1(x,\infty)$ in 
(\ref{rho}). In the first term contributing to spectral density, $K_1(x,x)$, the
microscopic limit and the integral can be interchanged. This establishes
universality of $K_1(x,x)$.

In the case of a Gaussian potential the edge contribution
can be obtained from the asymptotic expansion of  the Laguerre 
polynomials in this region (an expression in terms of Airy functions).
In a future publication, we hope to establish 
a possible relation with universal behavior 
of the $q_k^{2a}$ near the edge of the spectrum 
\cite{Bowick,KF}.

The above analysis carries through for the symplectic ensemble. In this case 
there are no contributions from the soft edge  and universality
of the microscopic spectral density can be shown rigorously.
For $\beta = 4$ (with an additional factor $1/2$ in 
the exponent of (\ref{zeig})),
the correlation functions can be expressed in terms of the kernel
\be
k_4(x,y) =\sum_{i,j=0}^{2n-1} Q_i(x) J_{ij} Q_j(y),
\label{k4}
\ee
where the $Q_i(x)$ are skew orthogonal polynomials of the first kind 
which are defined by the skew-scalar product 
\be
<f,g>_Q = \int_0^\infty \frac {dx}x e^{-2\phi_a(x)} f(x) (\hat L-\hat 1) g(x),
\ee
with the operator $\hat L$ defined in (\ref{L}). In this case we express
the $Q_i(x)$ in terms of the polynomials $q_k^{2a-1}(x)$,
\be
Q_k(x) = \sum_{l=0}^k S_{kl} q_l^{2a-1}(x).
\ee
The matrix elements of the operators are also in this basis.
From the orthogonality relation $ <Q_k,Q_l>_Q = J_{kl}$ it can be
shown that $S L S^T = - J$ from which we derive $S^T J S = XYX^{-1}$. 
Again, due
to the band structure of $L_{kl}$,  the 
range of the summations in this relation and in (\ref{k4}) differs by
a finite number of terms which can be neglected in the continuum limit.
We thus find
\be
k_4(x,y) \simeq 
\sum_{k,l=0}^{2n-1} q^{2a-1}_k(x) \left ( \hat Z \hat X^{-1} \right )_{kl} 
q_l^{2a-1}(y)\nonumber \\
= e^{\phi_a(y)}\int_0^\infty \frac {dz}z e^{-\phi_a(z)} \epsilon(y-z) 
K_{2n}^{2a-1}(x,z).
\ee
Universality of $k_4(x,y)$ thus follows from universality of 
$K_{2n}^{2a-1}(x,z)$. 
This relation is exact for a Gaussian potential and reproduces the
result found in \cite{Berbenni}.

In conclusion, 
we have shown that relations between the kernels for the chOE and chSE
and the kernel for chUE are not
accidental but follow from an intriguing underlying mathematical structure.
Under certain smoothness assumptions 
these relations are valid asymptotically for an arbitrary
polynomial potential. Microscopic universality for  $\beta =4 $ and 
in part for $\beta = 1$ thus follows
from universality at $\beta =2$ at hard edge of the spectrum.

This work was partially supported by the US DOE grant
DE-FG-88ER40388. P. Damgaard, B. Klein and S. Nishigaki are 
acknowledged for useful discussions.


\begin{references}

\bibitem{Porter}
C.E. Porter, `{\it Statistical theories of spectra: fluctuations}', Academic
Press, 1965; R. Haq, A. Pandey and O. Bohigas,
Phys. Rev. Lett. {\bf 48}, 1086 (1982).

\bibitem{hdgang}T. Guhr, A. M\"uller-Groeling and H.A. Weidenm\"uller,
%{\it Random Matrix Theories in quantum physics: Common concepts},
cond-mat/9707301, Phys. Rep. (in press).
    
\bibitem{AJM}J. Ambjorn, J. Jurkiewicz and Y. Makeenko,
Phys. Lett. {B251}, 517 (1990) ; J. Ambjorn and G. Akemann,
J. Phys. {A29}, L555 (1996); Nucl. Phys. {\bf B482}, 403 (1996).
\bibitem{Been}C.W.J. Beenakker, Nucl. Phys. {\bf B422}, 515 (1994).
\bibitem{bzgeneral}E. Br\'ezin and A. Zee, Nucl. Phys. {\bf B453}, 531 (1995).
\bibitem{Hack}G. Hackenbroich and H.A. Weidenm\"uller, Phys. Rev. Lett. {\bf
74} 4118 (1995).
\bibitem{Itoi}C. Itoi, cond-mat/9611214. 
\bibitem{Nishigaki-uni}S. Higuchi, C.Itoi,
S.M. Nishigaki and N. Sakai, Phys. Lett. {\bf B 398} (1997) 123.
%{\it Renormalization group approach to multiple
%arc random matrix models}, hep-th/9612237.
\bibitem{Brezinn}E. Br\'ezin and J. Zinn-Justin, Phys. lett. {\bf B288}, 
54 (1992).
\bibitem{brezin-hikami-zee}E. Br\'ezin, S. Hikami and A. Zee,
Nucl. Phys. {\bf B464}, 411 (1996).
\bibitem{Pzinn}P. Zinn-Justin, cond-mat/9705044;
Nucl. Phys. {\bf B497}, 725 (1997).
\bibitem{Zee-poly}E. Br\'ezin and A. Zee, Nucl. Phys. {\bf B402}, 613 (1993).
\bibitem{Deo}N. Deo, Nucl. Phys. {\bf B[FS]504}, 609 (1997); Nucl. Phys. B 464 
(1996) 463; N. Deo, S. Jain and B. S. Shastry Phys. Rev. E52, 4836 (1995).
%{\it Orthogonal polynomials
%and exact correlation functions for
%two cut random matrix models}, 
cond-mat/9703136.
\bibitem{GWu}T. Guhr and T. Wettig, 
%{\it Universal spectral correlations of
%the Dirac operator at finite temperature}, hep-th/9704055, 
Nucl. Phys. {\bf B506}, 589 (1997).
\bibitem{Sener1}A.D. Jackson, M.K. Sener and J.J.M. Verbaarschot, Nucl. Phys.
{\bf B479}, 707 (1996); Nucl. Phys.  {\bf B506}, 612 (1997).
\bibitem{Bowick}M.J. Bowick and E. Br\'ezin, Phys. Lett. {\bf B 268}, 21 
(1991).
\bibitem{KF}E. Kanzieper and V. Freilikher, Phys. Rev. Lett. {\bf 78},
3806 (1997); Phys. Rev. {\bf E 55}, 3712 (1997).
\bibitem{kanzieper}V. Freilikher, E. Kanzieper and I. Yurkevich, Phys. Rev.
{\bf E53}, 2200 (1996).
\bibitem{Eynard}B. Eynard, Nucl. Phys. {\bf B506}, 633 (1997). 
%{\it Eigenvalue distribution of large random
%matrices, from one matrix to several coupled matrices}, cond-mat/9707005.
\bibitem{Brezin-Hikami}E. Br\'ezin and S. Hikami, Phys. Rev. {\bf E56}, 264 
(1997). 
%{\it An extension of level spacing universality}, cond-mat/9702213.
\bibitem{Beenakker} C.W.J. Beenakker, Phys. Rev. Lett. {\bf 70}, 1155 (1993).
\bibitem{andreev}A.V. Andreev, O. Agam, B.D. Simons and B.L. Altshuler,
Nucl. Phys. {\bf B482} (1996) 536; A. Altland and M. Zirnbauer,
Phys. Rev. Lett. {\bf 77} (1996) 4536.
\bibitem{Akemann}G. Akemann,
P. Damgaard, U. Magnea and S. Nishigaki, Nucl. Phys. {\bf B 487[FS]}, 
721 (1997); S. Nishigaki, hep-th/9712051.
%{\it Universality of random matrices in the
%microscopic limit and the Dirac spectrum}, hep-th/9609174.
\bibitem{Akemannmulti}G. Akemann,
P.H. Damgaard, U. Magnea and S. Nishigaki, hep-th/9712006.
\bibitem{Damgaardmass}P.H. Damgaard and S. Nishigaki, hep-th/9711023; 
hep-th/9711096.
\bibitem{wilke}T. Wilke, T. Guhr and T. Wettig, hep-th/9711057.
\bibitem{Damgaard}P.H. Damgaard,  hep-th/9711110.
\bibitem{V2}J.J.M. Verbaarschot, Nucl. Phys. {\bf B426} (1994) 559.
\bibitem{Berbenni}M.E. Berbenni et al., hep-lat/9704018, Phys. Rev. Lett. (in 
press); hep-lat/9709102.
\bibitem{Slevin}
K. Slevin and T. Nagao, Phys. Rev. Lett. {\bf 70} (1993) 635.
\bibitem{martin}A. Altland and M.R. Zirnbauer, Phys. Rev. {\bf B55}, 1142 
(1997). 
\bibitem{BC} T. Banks and A. Casher,
Nucl. Phys. {\bf B169} (1980) 103.
\bibitem{SV}E.V. Shuryak and J.J.M. Verbaarschot,
Nucl. Phys. {\bf A560}, 306 (1993).
\bibitem{Ma} J.Z. Ma, T. Guhr and T. Wettig, hep-lat/9712026.
\bibitem{Markum}H. Markum et al., hep-lat/9709103
\bibitem{HV}M.A. Halasz and J.J.M. Verbaarschot,
Phys. Rev. Lett. {\bf 74}, 3920 (1995).
\bibitem{V} J. Verbaarschot, Phys. Rev. Lett. {\bf 72} (1994) 2531; Phys. Lett.
{\bf B329} (1994) 351; Nucl. Phys. {\bf B427} (1994) 434.
\bibitem{VZ}J. Verbaarschot and I. Zahed, Phys. Rev. Lett. {\bf 70}, 
3852 (1993).
\bibitem{Dyson_skew}F.J. Dyson, Comm. Math. Phys. {\bf 19} 235, (1970).
\bibitem{mahoux}G. Mahoux and M. Mehta, Comm. Math. Phys. {\bf 79}, 327
(1981); Indian J. Pure Appl. Math. {\bf 22}, 531 (1991); J. Phys. {\bf A14},
579 (1981); J. Phys. France {\bf I} (1991) 1093.
\bibitem{BN}E. Br\'ezin and H. Neuberger, Nucl. Phys. {\bf B350}, 513 (1991).
\bibitem{Nagao}T. Nagao and M. Wadati, J. Phys. Soc. Jpn. {\bf 60}, 3298
(1991); {\bf 61}, 78 (1992); {\bf 61}, 1910 (1992).
\bibitem{Bronk}
B. Bronk, J. Math. Phys. {\bf 6} (1965) 228.
\bibitem{Kahn}
D. Fox and P. Kahn, Phys. Rev. {\bf 134} (1964) B1151.
\bibitem{Widom}
C. Tracy and H. Widom,  Comm. Math. Phys. {\bf 161} (1994) 289.
\bibitem{Nagao-Sl}T. Nagao and K. Slevin, J. Math. Phys. {\bf 34} (1993) 2075;
J. Math. Phys. {\bf 34} (1993) 2317.
\bibitem{Forrester-Nagao}
T. Nagao and P.J. Forrester, Nucl. Phys. B435 (1995) 401-420.
\end{references}
\end{document}